\documentclass[epj,referee]{svjour}

\usepackage{graphics}

\begin{document}

\title{Resonance phenomena in discrete systems with bichromatic input signal}
\author{K.P.Harikrishnan\inst{1} \thanks{Corresponding author:email: 
$kp_{\_}hk2002@yahoo.co.in$}  and  G.Ambika\inst{2}
}
\institute{Department of Physics, The Cochin College, Cochin-682002, 
India \and Indian Institute of Science Education and Research, Pune-411008,  
India}
\abstract{
We undertake a detailed numerical study of the twin phenomenon of stochastic 
and vibrational resonance in a discrete model system in the presence of bichromatic 
input signal. A two parameter cubic map is used as the model that 
combines the features of both bistable and  threshold settings. Our analysis  
brings out several 
interesting results, such as, the existence of a cross over behavior from vibrational 
to stochasic resonance and the possibility of using stochastic resonance as a 
filter for the selective detection/transmission of the component frequencies in a 
composite signal. The study also reveals a fundamental difference between the 
bistable and threshold mechanisms with respect to amplification of a 
multi signal input. 
}
\maketitle

\section{Introduction}
\label{sec:1}
During the last two decades, investigation of signal processing in nonlinear 
systems in the presence of noise has revealed several interesting phenomena, 
the most important being \emph{stochastic resonance}(SR)[1]. It can roughly be 
considered as the optimisation of certain system performance by noise. The 
interest in the study of nonlinear noisy systems has increased due to the 
applications in the modelling of a great variety of phenomena of relevance in 
physics,chemistry and lifesciences[1-3]. 

In SR, the response of a 
nonlinear system to a weak input signal is significantly increased with 
appropriate tuning of the noise intensity[1]. When a subthreshold signal 
$I(t)$ is input to a nonlinear system $g$ together with a noise 
$\zeta (t)$, if the filtered output $O(t) \equiv g(I(t)+ \zeta (t))$ 
shows enhanced response that contains the information content of $I(t)$, 
then SR is said to be realised in the system. The mechanism first used by 
Benzi, Nicolis etc [4,5] to explain natural phenomena is now being used 
for a large variety of interesting applications like modelling biological 
and ecological 
systems [6], lossless communication purposes etc [7]. Apart from these, it 
has opened up a vista of many related resonance phenomena [8] which are 
equally challenging from the point of view of intense research. The most 
important among these, which closely resembles SR, is the \emph {vibrational resonance} 
(VR) [9], where a high frequency forcing plays the role of noise and 
amplify the response to a low frequency signal in bistable systems. 
In VR, analogous to SR, the system response shows a bell shaped resonant form as a 
function of the amplitude of the high frequency signal. In this work, we try to 
capture numerically some interesting and novel aspects of SR and VR, using a simple 
discrete model, namely, a two parameter bimodal cubic map.

In the early stages of the development of SR, most of the studies were done 
using a dynamical setup with bistability, modelled by a double well 
potential. Here SR is realised due to the shuttling between the two stable 
states at the frequency of the subthreshold signal with the help of noise. 
Thus if the potential is 
\begin{equation}
V(X) = -a X^2/2 + b X^4/4
\label{eqn1}
\end{equation}
in the presence of a signal and noise, the dynamics can be modelled by an 
overdamped oscillator
\begin{equation}
\dot{X} = a X - b X^3 + ZSin \omega t + E \zeta(t)
\label{eqn2}
\end{equation}
If $C_{th}$ is the threshold at which deterministic switching (with noise 
amplitude $E=0$) is 
possible, then well to well switching due to SR occurs when
\begin{equation}
-C_{th} > (ZSin \omega t+E \zeta(t)) > C_{th}
\label{eqn3}
\end{equation}
that is, twice in one period of the signal $T = 2 \pi/\omega$.

The characterisation of SR in this case is most commonly done by computing 
the signal to noise ratio (SNR) from the power spectrum of the output as
\begin{equation}
SNR = 10log_{10} (S/N)  dB
\label{eqn4}
\end{equation}
where $N$ is the average background noise around the signal $S$.
If SR occurs in the system, then the SNR goes through a maximum giving a bell 
shaped curve as $E$ is tuned.

SR has also been observed in systems with a single stable state with an 
escape scenerio. These \emph{threshold} systems, in the simplest case, can 
be modelled by a piecewise linear system or step function

$$g(u) = -1 \hspace {12pt} u < C_{th}$$\\
$$\hspace {22pt} = +1 \hspace {12pt} u > C_{th}$$\\
The escape with the help of noise is followed by resetting to the monostable 
state. In this case, a quantitative characterisation is possible directly 
from the output, but only in terms of probabilities. If $t_n$ are the escape 
times, the inter spike interval (ISI) is defined as $T_n = t_{n+1}-t_n$ and 
$m(T_n)$ is the number of times the same $T_n$ occurs. For SR to be realised 
in the system, the probability $p_n = m(T_n)/N$ ($N$ is the total number of 
escapes) has to be maximum corresponding to the signal period $T$ at an 
optimum noise amplitude.

\begin{figure}
\resizebox{1.0\columnwidth}{!}{\includegraphics{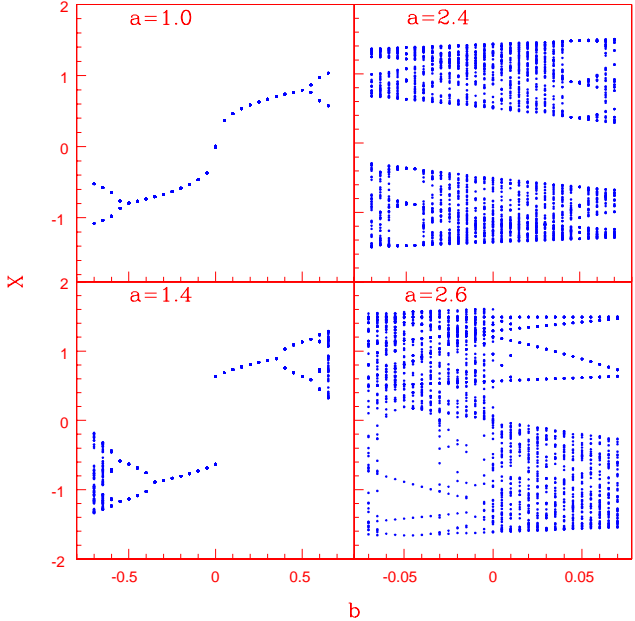}}
\caption{The bifurcation structure of the cubic map given in eq.(5) corresponding 
to four values of the parameter $a$, to show the 
bistability in periodic  and chaotic attractors  with the basin 
boundary at $X=0$. It is clear that the bistability just begins at $a = 1.0$ and 
ends at $a = 2.6$. Note that the range of $b$ values for bistability is different 
for periodic and chaotic attractors.} 
\label{Fig.1label}
\end{figure}

In the case of VR, the high frequency forcing takes the role of noise to boost 
the subthreshold signal. The system is then under the action of a two 
frequency signal, one of low frequency and the other of high frequency, with or 
without the presence of noise. 
Most of the studies in VR also have been done using the standard model of 
overdamped bistable oscillators [10]. But recently, VR has been shown to 
occur in a spatially extended system of coupled noisy oscillators [11] and two 
coupled anharmonic oscillators [12] in the bistable setup. In the threshold setup, 
VR has been shown to occur only in one system, namely, in the numerical simulation 
of the FitzHugh-Nagumo (FHN) 
model in the excitable regime along with the experimental confirmation using an 
electronic circuit [13]. Here we show for the first time the occurance of VR 
in a \emph{discrete} system, both in bistable and threshold setups.

There are situations where SR and VR are to be optimised by adapting to or 
designing the dynamics of the system. This is especially relevant in natural 
systems  where the noise is mostly from the 
environmental background and therefore not viable to fine tuning. Similarly, 
depending on the context or application, the nature of the signal can also 
be different, such as, periodic, aperiodic, digital, composite etc. The 
classical SR deals with the detection of a single subthreshold signal 
immersed in noise. However, in many practical situations, a composite signal 
consisting of two harmonic components in the presence of 
background noise is encountered, as for example, in biological systems for the  
study of planktons and human visual cortex [14], in laser physics [15] and in 
accoustics [16]. Studies involving such \emph{bichromatic} signals are also 
relevant in communication, since one can address the question of the carrier signal 
itself amplifying the modulating signal.
Moreover, two frequency signals are commonly used in 
multichannel optical communication systems based on wave length division 
multiplexing (WDM) [17]. But only very few studies of SR have been carried out 
using  the bichromatic signals to date [18-21], each of them 
pertaining to some specific dynamical setups and with continuous systems. 
This is one of the motivations for us to 
undertake a detailed numerical analysis of SR and VR with such signals using a 
discrete model. 
It is a discretised version of the 
overdamped bistable oscillator, but with the added feature of an inherent 
escape mode also. Hence it can function in both setups, bistable and 
threshold, as a stochastic resonator. 

There are many situations, especially in 
the biological context, such as, host-parasite model, virus-immune model etc., where 
discrete systems model the time developments directly. They can behave differently,  
especially in the presence of high frequency modulation and background noise.
The benefit of high 
frequency forcing has been studied in the response of several biological 
phenomena[22]. High frequency stimulation treatments in Parkinson's disease and 
other disorders in neuronal activity have also been reported[23]. Moreover, it is 
also known that optimum high frequency modulation improves processing of low 
frequency signal even in systems without bistability where noise can induce 
the required structure[11]. Thus a study of resonance phenomena in discrete 
systems can lead to qualitatively different results having potential practical 
applications. We do observe some novel features which have not been reported 
so far for continuous systems.

Our paper is organised as follows: In \S 2, the model system used for the analysis 
is introduced. In \S 3, we study SR in the model numerically with bichromatic 
signal treating it as a bistable system. Numerical and analytical results of VR 
in the bistable setup are presented in \S 4. \S 5 discusses SR and VR with the cubic 
map as a threshold system. Results and discussions are given given in \S 6.

\begin{figure}
\resizebox{1.0\columnwidth}{!}{\includegraphics{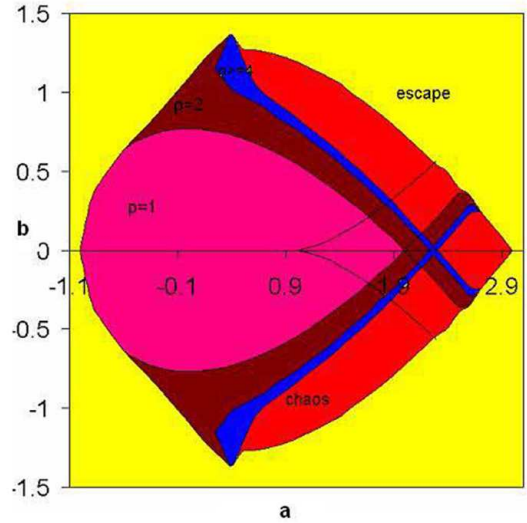}}
\caption{Parameter space plot in $(a,b)$ plane for the cubic map given by eq.(5). 
Regions of periodic attractors with period 1(p=1), period 2(p=2) etc. and that of 
chaotic attractors and escape are shown in different colours. Region within the 
quadrilaterals $(a > 1.0)$ corresponding to $p=1, p=2$ and chaos represent 
the respective bistable regions.}
\label{Fig.2label}
\end{figure}

\section{The model system}
\label{sec:2}
The model used for our analysis is a two parameter cubic map given by
\begin{equation}
X_{n+1}= f(X_{n})= b + a X_{n} - X_{n}^3
\label{eqn5}
\end{equation}
It is a discrete version of the Duffing's double well potential and is the 
simplest possible nonlinear discrete system that combines the desirable 
features of  bistable and threshold setups. Hence it is extremely useful in 
the study of resonance phenomena caused by noise or high frequency signal. 
Similar systems in the continuous cases include the FHN model[23] for neuronal 
firing with two widely different time scales.
 
The system has been studied in great detail both analytically and numerically 
and has been shown to possess a 
rich variety of dynamical properties including bistability [24]. In particular,
if $a_1$ is the value of the parameter at which $f^{\prime}(X_i,a_1,b) =1$,
then for 
$a>a_1$, there is a window in $b$, where bistability is observed. The bistable 
attractors are clearly separated with $X>0$ being the basin of one and 
$X<0$ that of the other. It is found that for $a$ in the range $1.0 < a < 2.6$, 
the system has periodic states and chaotic states. As $a$ increases from $1.0$, 
the periodicity of the bistable attractors keep on doubling while the width of 
the window around $b$ decreases correspondingly. For $a=2.4$, two chaotic attractors 
co-exist in a very narrow window around $b$. For $a<1.0$, the system has a 
monostable period 1 state and for $a>2.6$, there is merging of the chaotic states 
followed by escape. All these can be clearly seen from Fig.1 where bifurcation 
structure of the cubic map corresponding to four $a$ values are shown. A detailed 
stability analysis fixes the different asymptotic behavior in its parameter plane 
$(a,b)$ as shown in Fig.2. Regions of periodicities $(p=1, p=2)$ etc., chaos and 
escape in the parameter plane can be clearly seen. The quadrilateral regions in the 
area marked $p=1,p=2$ and chaos represent bistability in the respective attractors. 

The system when driven by a gaussian noise and a periodic signal becomes 
\begin{equation}
X_{n+1} = b + a X_n - X_{n}^3 + E \zeta(n) + Z F(n)
\label{eqn6}
\end{equation}
where we choose $\zeta(n)$ to be a gaussian noise time series with zero mean and 
$F(n)$ is the 
periodic signal sampled in unit time step. The amplitude of the noise and the 
signal can be varied by 
tuning E and Z respectively. It can be shown that in the regime of chaotic 
bistable attractors $(a=2.4,b=0.01)$, a subthreshold input signal can be 
detected using the inherent chaos in 
the system without any external noise(E=0). Taking the signal 
$F(n) = Z Sin(2 \pi \nu n)$, with $Z = 0.16$, the system shows SR type 
behavior 
for an optimum range of frequencies as shown in Fig.3, where the output SNR 
is plotted 
as a function of the frequency $\nu$. It implies that a subthreshold signal  
can be detected by passing through a bistable system making use of the 
inherent chaos in it without the help of any external noise. This 
phenomenon is known as \emph {deterministic resonance}. In the regime of 
periodic bistable attractors $(a=1.4,b=0.01)$ with a single subthreshold 
signal, the system shows conventional SR as well as chaotic resonance (CR), 
and using this model, we 
have recently reported some new results including enhancement of SNR via 
coupling [25].

\begin{figure}
\resizebox{1.0\columnwidth}{!}{\includegraphics{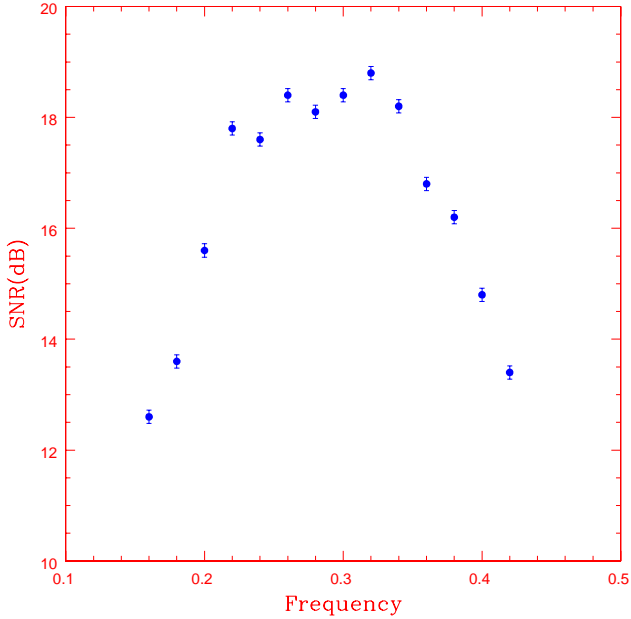}}
\caption{Variation of SNR with frequency for the bistable cubic map in the 
chaotic regime $(a=2.4)$ with the noise amplitude $E=0$. Note that a 
subthreshold signal $(Z=0.16)$ can be detected for an intermediate range of 
frequencies, without any external noise.}
\label{Fig.3label}
\end{figure}

\section{SR with bichromatic input signal}
\label{sec:3}
In this section, we undertake a numerical study of SR using the cubic map in the 
bistable regime. The input signal $F(n)$ is taken as a superposition of at least 
two fundamental frequencies, $\nu_1$ and $\nu_2$. Before going into the numerical 
results, we briefly mention some already known results for multisignal inputs in the 
case of over damped bistable potential. 

The most popular theory for the analytical description of SR is the linear 
response theory(LRT)[26-29].
According to this theory, the response of a nonlinear stochastic system 
$X(t)$ to a weak external force $F(t)$ in the asymptotic limit of large 
times is determined by the integral relation [26]
\begin{equation}
X(t) = <X_0> + \int_{-\infty}^{\infty} R(t-\tau) F(\tau) d \tau
\label{eqn7}
\end{equation}
where $<X_0>$ is the mean value of the state variable for $F(t)=0$. 
Without lack of generality, one can set $<X_0>=0$. The function $R(t)$ in 
(7) is called the response function. For a harmonic signal, the system response 
can be expressed through the function $R(\omega)$ which is the Fourier 
transform of the response function:
\begin{equation}
X(t) = Z |R(\omega)| Sin(2 \pi \nu t + \psi)
\label{eqn8}
\end{equation}
where $\psi$ is a phase shift.

The LRT can be naturally extended to the case of multifrequency signals. Let 
the signal $F(t)$ be a composite signal of the form:
\begin{equation}
F(t) = Z \sum_{k=1}^{n} Sin(2 \pi \nu_{k}t)
\label{eqn9}
\end{equation}
where $\nu_{k}$ are the frequencies of the discrete spectral components with 
the same amplitude $Z$. Then according to LRT, the system response can be 
shown to be [29]
\begin{equation}
X(t) = Z \sum_{k=1}^{n} |R(\omega_{k})| Sin(2 \pi \nu_{k}t + \psi_{k})
\label{eqn10}
\end{equation}
which contains the same spectral components at the input. We now show 
numerically that the same is true for a 
discrete system as well, in the bistable setup.

For the remaining part of the paper, we fix $a=1.4$ and $b=0.01$ in the region 
of bistable periodic attractors and $\zeta(n)$ is a gaussian time series whose 
amplitude can be tuned by changing the value of $E$. The system is driven by a 
composite signal consisting of a combination of two frequencies $\nu_1$ 
$\nu_2$, given by
\begin{equation}
Z F(n) = Z(Sin 2 \pi \nu_1 n + Sin 2 \pi \nu_2 n)
\label{eqn11}
\end{equation}
The value of $Z$ is fixed at $0.16$ so that the signal is well below threshold.
It should be noted that because of the iteration with unit time step
$(ie, n=1,2,3,...)$, the available range of frequencies is limited to 
$\nu_k \equiv (0, 0.5)$. We use different combination of frequencies 
$(\nu_1,\nu_2)$ for numerical simulation in the presence of noise by tuning 
the value of $E$. For convenience, $\nu_1$ is fixed at $0.125$ and $\nu_2$ is 
varied from $0.02$ to $0.5$ in steps of $0.01$. For each selected combination, 
the output power spectrum is calculated using the FFT algorithm for different 
values of $E$. A typical power spectrum for 
$(\nu_1,\nu_2) = (0.125,0.05)$ is shown in Fig.4(a). The procedure is 
repeated with $F(n)$ consisting of a combination of 3 frequencies and a 
power spectrum for $(\nu_1,\nu_2,\nu_3) = (0.125,0.075,0.2)$ is shown in 
Fig.4(b). To compute the power spectrum, only the inter-well transitions are 
taken into account and all the intra-well fluctuations are suppressed with a 
two state filtering. It is clear that, only the fundamental frequencies 
present in the input are enhanced.

\begin{figure}
\resizebox{1.0\columnwidth}{!}{\includegraphics{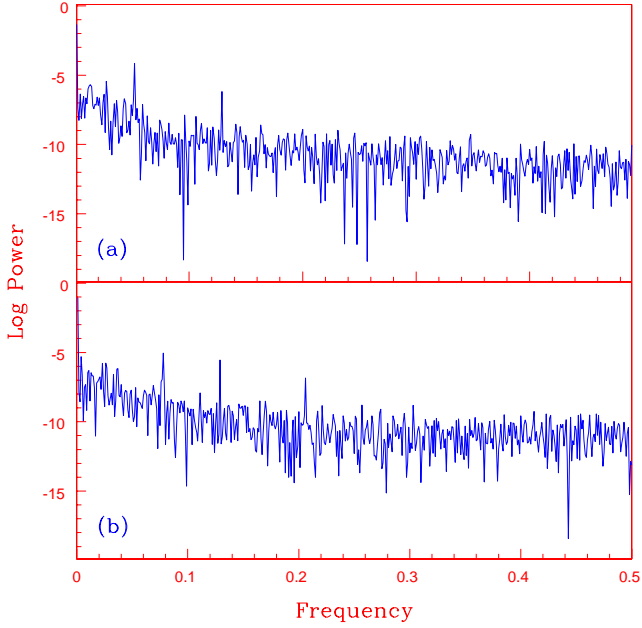}}
\caption{Power spectrum for the bistable system (6) with a composite signal 
consisting of $(a)$ 2 frequencies $(\nu_1,\nu_2) = (0.125,0.05)$ and $(b)$ 3 
frequencies $(\nu_1,\nu_2,\nu_3) = (0.125,0.075,0.2)$, with $Z=0.16$ and 
$E=0.4$ in both cases. The system parameters are $a=1.4$ and $b=0.01$.}
\label{Fig.4label}
\end{figure}

\begin{figure}
\resizebox{1.0\columnwidth}{!}{\includegraphics{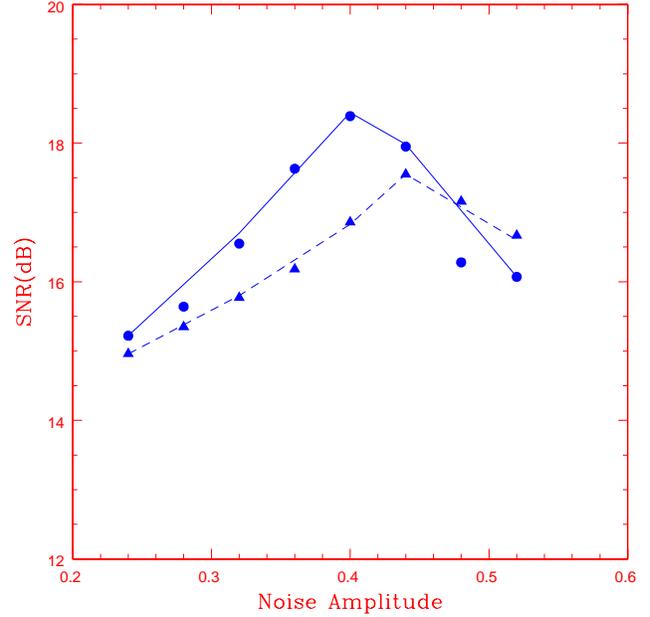}}
\caption{Variation of SNR with noise amplitude for the 2 frequencies shown in 
$Fig.4(a)$, with the filled circle connected by solid line representing the 
higher frequency $\nu_1$.}
\label{Fig.5label}
\end{figure}

\begin{figure}
\resizebox{1.0\columnwidth}{!}{\includegraphics{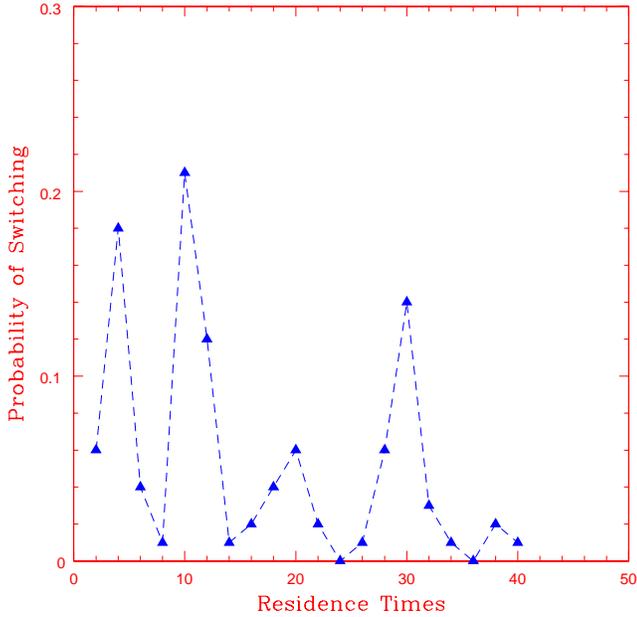}}
\caption{The RTDF for the system (6), with a composite signal consisting of 2 
frequencies $\nu_1=0.125$ and $\nu_2=0.05$ and amplitude $Z=0.16$. The noise 
amplitude is put at the optimum value $0.4$. Note that the peaks are 
synchronised with half the periods corresponding to the input frequencies.}
\label{Fig.6label}
\end{figure}

We now concentrate on a combination of 2 frequencies and compute the 
two important quantifiers of SR, namely, the SNR and the Residence Time 
Distribution Function(RTDF). For the frequencies in Fig.4(a), the SNR is 
computed from the power spectrum using equation (4)
for a range of values of $E$ and the results are shown in Fig.5. 
The RTDF measures the probability distribution of the average times the 
system resides in one basin, as a function of different periods. If T is the 
period of the applied signal, the distribution will have peaks corresponding 
to times $(2n+1)T/2$, n=0,1,2...For the system (6) with $(\nu_{1},\nu_{2}) = 
(0.125,0.05)$, the results are shown in Fig.6. Note that there are only peaks 
corresponding to the half integer periods of the two applied frequencies.

The above computations are repeated taking various combinations of 
frequencies $(\nu_1,\nu_2)$, both commensurate and non-commensurate.  
For a fixed combination of $(\nu_{1},\nu_{2})$, the calculations  
are also done by changing the signal amplitude $Z$ of one and both signals. 
Always the results remain qualitatively the same and only the fundamental 
frequencies 
present in the input are amplified at the output. If $Z$ becomes very small
$(< 0.1)$ compared to the noise level, then the phenomenon of SR disappears 
altogether and the signal remains undetected in the background noise.

In all the above computations, we used \emph{additive} noise, where the noise 
has been added to the system externally. But in many natural systems, noise 
enters through an interaction of the system with the surroundings, that is,
through a parameter modulation, rather than a simple addition. Such a 
\emph{multiplicative} noise occurs in a variety of physical phenomena [30] 
and can, in principle, show qualitatively different behavior in the presence 
of a periodic field [31]. To study its effect on the bistable 
system, equation (6) is modified as
\begin{equation}
X_{n+1} = b + a(1+E \zeta (n))X_{n} - X_{n}^3 + Z F(n)
\label{eqn12}
\end{equation}
The noise is added through the parameter $a$ which determines the nature of 
the bistable attractors. With $a=1.4$ and $b=0.01$, the system is now driven 
by a signal of single frequency $\nu_{1}=0.125$ and a multisignal with 2 
frequencies $(\nu_{1},\nu_{2})$, with $Z=0.16$. The power spectrum for single 
frequency and multiple frequencies are shown in Fig.7(a) and (b) 
respectively. The corresponding SNR variation with noise $E$ are shown in 
Fig.8(a) and (b). Note that the results are qualitatively identical to 
that of additive noise, but the optimum SNR and the corresponding noise 
amplitude are comparitively much higher in this case. Thus 
our numerical results indicate that a bistable system responds only to the 
fundamental frequencies in a composite signal and not to any mixed modes.  

\begin{figure}
\resizebox{1.0\columnwidth}{!}{\includegraphics{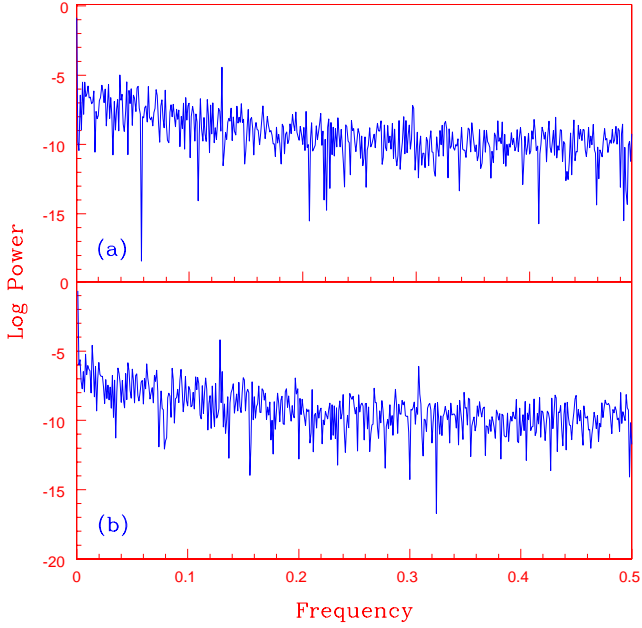}}
\caption{The power spectrum for the bistable system (12), with the signal 
$F(n)$ consisting of $(a)$ one frequency $\nu_1=0.125$ and $(b)$ 2 frequencies 
$\nu_1=0.125$ and $\nu_2=0.3$. The parameter values used are $Z=0.16, E=1.0, 
a=1.4$ and $b=0.01$.}
\label{Fig.7label}
\end{figure}

\begin{figure}
\resizebox{1.0\columnwidth}{!}{\includegraphics{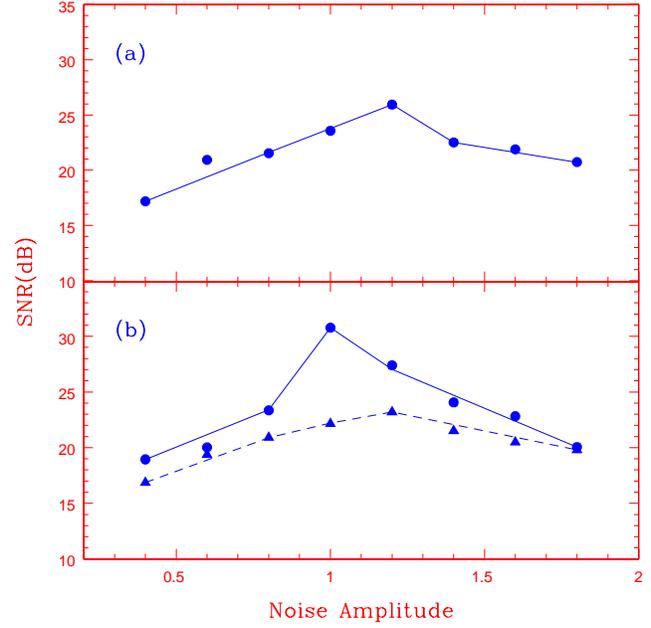}}
\caption{Variation of SNR with noise for the system(12), for $(a)$ single 
frequency $\nu_1=0.125$ and $(b)$ 2 frequencies $\nu_1=0.125$ (filled 
circles) and $\nu_2=0.3$. Note that the optimum SNR of $\nu_1$ increases by 
about 5 dB when a second signal $\nu_2$ is added.}
\label{Fig.8label}
\end{figure}

\begin{figure}
\resizebox{1.0\columnwidth}{!}{\includegraphics{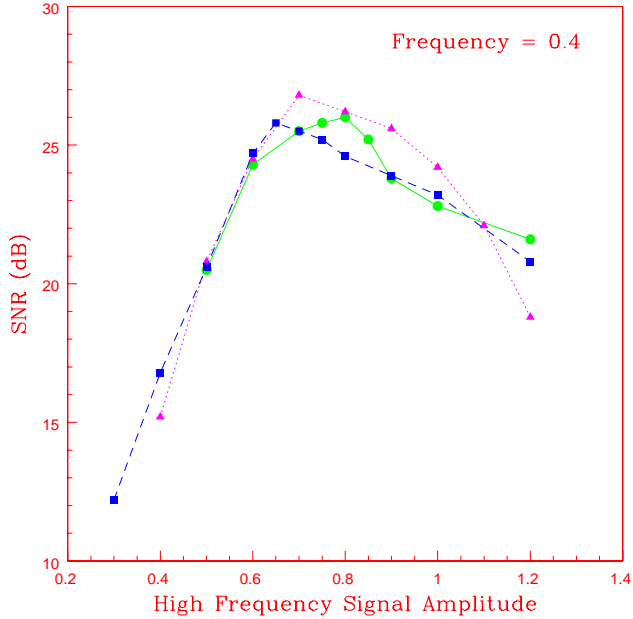}}
\caption{Variation of SNR corresponding to the frequency $\nu_1 = 0.125$ as a 
function of the amplitude $Z(2)$ of the high frequency signal with $\nu_2 = 0.4$,  
showing VR in the bistable cubic map with a bichromatic input signal given by 
eq.(13). The three curves correspond to different values of noise amplitude, 
namely, $E=0.06$ (circle connected by solid line), $E=0.12$ (triangle connected by 
dotted line) and $E=0.16$ (squares conneted by dashed line). The optimum values 
of $Z(2)$ shifts towards the left as $E$ increases.} 
\label{Fig.9label}
\end{figure}

\section{VR: Numerical and analytical results}
\label{sec:4}
In order to study VR in the system, the input periodic driving is 
modified as 
\begin{equation}
Z F(n) = Z(1) Sin(2 \pi \nu_{1}n) + Z(2) Sin(2 \pi \nu_{2}n)
\label{eqn13}
\end{equation}
where $\nu_1$ is the low frequency signal which is fixed at $0.125$ with its amplitude 
$Z(1)$ at the subthreshold level $0.16$. It is added with a signal of high frequency 
$\nu_2 (> \nu_1)$ whose amplitude $Z(2)$ is tuned to get VR. We have used a number of  
values for 
$\nu_2$ over a wide range from $0.15$ to $0.5$ to study VR in the system. A small 
amount of noise is also added as in eq.(6) which represents tiny random 
fluctuations present in all practical systems.

\begin{figure}
\resizebox{1.0\columnwidth}{!}{\includegraphics{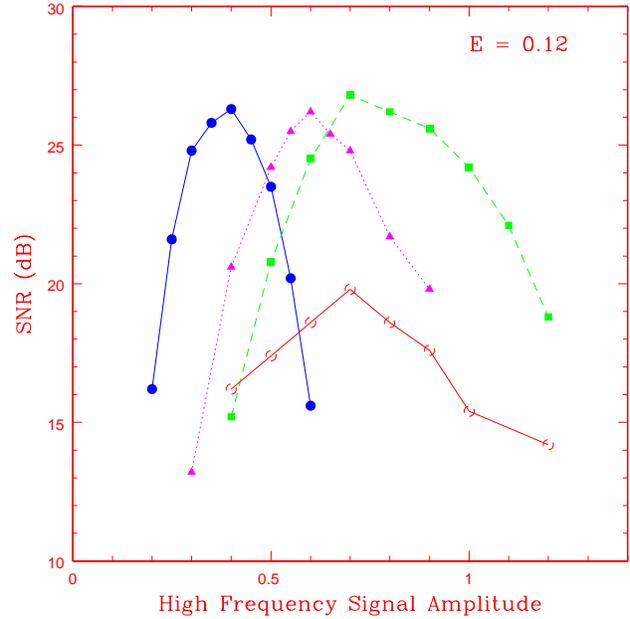}}
\caption{Variation of SNR for $\nu_1$ as a function of $Z(2)$ corresponding to 
four values of $\nu_2$ which are $0.18$ (filled circle connected by solid line), 
$0.32$ (triangles connected by dotted line), $0.40$ (squares connected by dashed 
line) and $0.46$ (open circles connected by solid line). As $\nu_2 >> \nu_1$, it 
becomes less effective in producing VR as indicated by the reduction in the 
optimum value of SNR.} 
\label{Fig.10label}
\end{figure}

By taking $\nu_2=0.4$ and $E=0.12$, the system is iterated by tuning the high 
frequency signal amplitude. The variation of SNR for the signal computed from 
the output time series as a function of $Z(2)$, clearly indicates VR in the 
system. The procedure is repeated by changing $E$ to study the influence of 
noise level on VR. The results are shown in Fig.9 for 3 values of $E$. It is 
found that as $E$ decreases, the optimum SNR shifts towards the higher value of 
$Z(2)$. This result has been reported earlier in the case of continuous 
systems also[10,12].

The computations are repeated by taking different values of $\nu_2$ in the 
range $0.15$ to $0.5$ by fixing $E=0.12$. The SNR as a function of $Z(2)$ for 
four values of $\nu_2$ are shown in Fig.10. Two results are evident from the 
figure. The optimum SNR is independant of the high frequency when 
$(\nu_2 - \nu_1)$ is small. But for $\nu_2 >> \nu_1$, the optimum SNR decreases 
appreciably. This implies that the effectiveness of the high frequency in 
producing VR is reduced as it becomes too large compared to the low frequency 
signal. It can also be seen that the optimum value of $Z(2)$ depends on $\nu_2$. 
Infact, it increases with $\nu_2$. Thus, a higher value of $\nu_2$ requires a 
larger amplitude and becomes less effective as $\nu_2 >> \nu_1$.

One reason for this is that, due to the unit time sampling of the signals, the 
fluctuation in the signal values will be more as the frequency increases. Hence 
the average will be closer to zero, higher the frequency, which makes it necessary 
to have larger amplitude. But once the optimum amplitude becomes too large, 
escape becomes possible in all time scales making it less effective. It can also 
be shown that the high frequency forcing can change the effective value of the 
parameter $a$ so that the map can come out of the bistable window. This is 
discussed in more detail below.

The dependance of optimum $Z(2)$ on $\nu_2$ also implies that the system shows 
the so called \emph{bona fide} resonance[32,33]. This is shown in Fig.11, where 
the SNR is plotted as a function of the high frequency for two values 
$(0.5 \& 0.7)$ of $Z(2)$, with $E$ fixed at $0.12$.

We now briefly discuss some analytical results for VR in the cubic map and show that 
the high frequency forcing can change the effective value of the parameter $a$. 
Here we consider the case $\nu_2 >> \nu_1$, in the noise free limit, $E=0$. In 
parallel with the theory developed to explain VR in continuous systems[34], we 
analyse the effect of the widely differing frequencies for the system under study: 
\begin{equation}
X_{n+1} = b + a X_{n} - X_{n}^3 + A Sin \omega_{1} n + B Sin \omega_{2} n
\label{eqn14}
\end{equation}
where $\omega_{1,2} = 2 \pi \nu_{1,2}$. Taking $A=0$, we look for a solution[34]
\begin{equation}
X_n = Y_n - {{B Sin \omega_2 n} \over {\omega_{2}^2}}
\label{eqn15}
\end{equation}
While the first term $Y_n$ varies significantly only over time of the order $n$, 
the second term varies rapidly within an iteration and hence can be averaged. 
Putting (15) in (14) and averaging, we get:
\begin{equation}
Y_{n+1} = b + a Y_n - Y_{n}^3 - {{3 Y_n B^2} \over {2 \omega_{2}^2}}
\label{eqn16}
\end{equation}
That is,
\begin{equation}
Y_{n+1} = b + a^* Y_n - Y_{n}^3
\label{eqn17}
\end{equation}
where
\begin{equation}
a^* = a - {{3 B^2} \over {2 \omega_{2}^2}}
\label{eqn18}
\end{equation}
Thus the effect of the high frequency forcing is to reset the parameter $a$ as 
$a^*$. Hence only for the choice of $B$ and $\omega_2$ that retains $a^*$ in the 
bistable window $(1.0 < a^* < 2.6)$, do we expect shuttling behavior at the 
low frequency signal for $A \neq 0$. Now the lower limit for $a$ becomes
\begin{equation}
a \geq 1.0 + {{3 B^2} \over {2 \omega_{2}^2}}
\label{eqn19}
\end{equation}
which gives
\begin{equation}
({{B} \over {\omega_{2}^2}})^2 \leq {{{2} \over {3}} (a - 1.0)}
\label{eqn20}
\end{equation}
Thus, for a given choice of $a$, there is an upper limit for 
$({{B} \over {\omega_{2}}})$ for retaining the bistability in the system. The 
actual values may get modified in the presence of noise. Also, this mechanism 
provides two parameters $B$ and $\omega_2$ that can be tuned in a mutually 
compramisable manner to obtain VR or even SR with added noise. Moreover, even 
when $a > 2.6$, $a^*$ can be $< 2.6$ and hence this increases the virtual 
window of bistability in the system providing greater range for applicability.

\section{SR and VR in the threshold setup}
\label{sec:5}
As said earlier, the domains of the bistable attractors in the cubic map are 
clearly separated with the boundary $X=0$. Hence the cubic map can also be 
considered as a nondynamical threshold system with a single stable state 
having a potential barrier. Here the system generates an output \emph{spike} 
only when the combined effort of the signal and the noise pushes it across 
the potential barrier $(at X=0)$ in one direction:
\begin{equation}
[E \zeta (n) + Z F(n)] > C_{th}
\label{eqn21}
\end{equation} 
It is then externally reinjected back into the basin. The output thus 
consists of a series of spikes similar to a random telegraph process. The 
study of SR and VR in such systems assumes importance in the context of biological 
applications and in particular the integrate and fire models of neurons 
where SR and VR have been firmly established[3,35].

The computations are done using equation (6), initially with a single frequency 
signal, $F (n) = Sin(2 \pi \nu_{1} n)$. We start from an initial condition in 
the negative basin and when the output crosses the threshold, $X=0$, it is 
reinjected back into the basin by resetting the initial conditions. This is 
repeated for a sufficiently large number of escapes and the ISIs are 
calculated. The ISIs are then normalised in terms of the 
periods $T_{n}$ and the probability of escape corresponding to each $T_{n}$ 
is calculated as the ratio of the number of times $T_{n}$ occurs to the 
total number of escapes. The whole procedure is repeated tuning the noise 
amplitude $E$. It is found that the ISI is synchronised with the period of the 
forcing signal for an \emph{optimum} noise amplitude (Fig.12), indicating SR 
for the frequency $\nu_{1}$.

\begin{figure}
\resizebox{1.0\columnwidth}{!}{\includegraphics{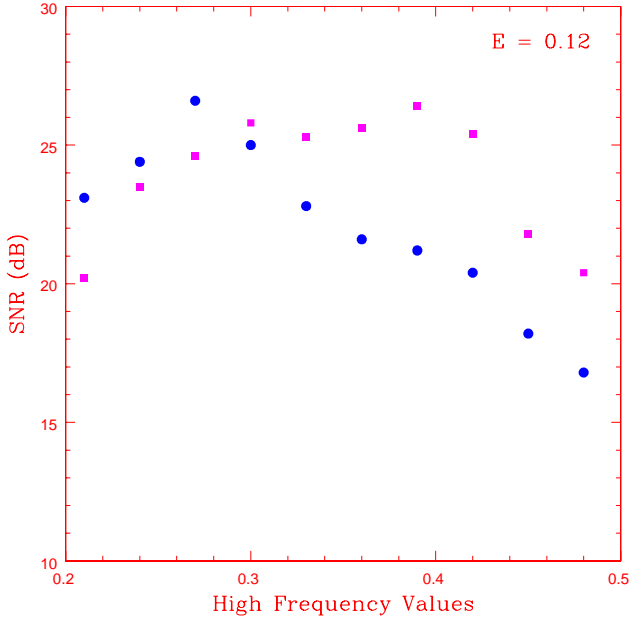}}
\caption{Evidence for bona-fide resonance in the cubic map. Variation of SNR 
corresponding to $\nu_1=0.125$ as function of the high frequency $\nu_2$ for two 
values of its amplitude, namely, $Z(2)=0.5$ (circles) and $Z(2)=0.7$ (squares). 
The amplitude of $\nu_1$ is fixed at $0.16$ and $E=0.12$.} 
\label{Fig.11label}
\end{figure}

\begin{figure}
\resizebox{1.0\columnwidth}{!}{\includegraphics{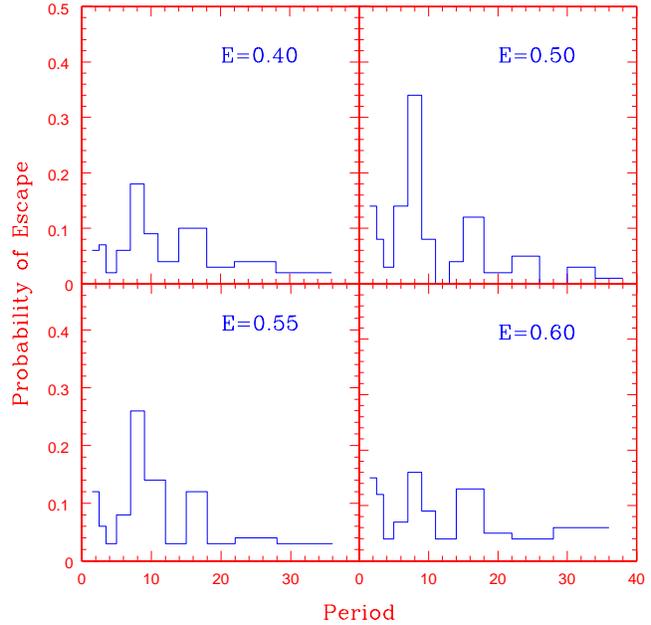}}
\caption{The probability of escape for the system (6) as a function of different 
periods, when it is used as a threshold system with the signal $F(n)$ having a 
single frequency $\nu_1 = 0.125$ and a subthreshold amplitude $Z=0.16$. 
The escape probability is plotted corresponding to 
four different noise amplitudes $E$. Note that the probability corresponding to the 
signal period $(T=8)$ passes through an optimum value for $E=0.5$.}
\label{Fig.12label}
\end{figure}

The calculations are now repeated by adding a second signal of frequency 
$\nu_{2}$ and amplitude same as that of $\nu_{1}$. Again different values of 
$\nu_{2}$ in the range $0.02$ to $0.5$ are used for the calculation. It is then 
found that apart from the input frequencies $\nu_{1}$ and $\nu_{2}$, a third 
frequency, which is a mixed mode is also enhanced at the output, at a lesser 
value of the noise amplitude. To make it clear, the amplitude $Z$ of both 
$\nu_{1}$ and $\nu_{2}$ are reduced from $0.16$ to $0.08$, so that they become 
too weak to get amplified. The results of computations are shown in Fig.13 and 
Fig.14, for a combination of input signals $(\nu_1,\nu_2)=(0.125,0.033)$.
Fig.13 represents the probability of escape corresponding to different periods,
for the optimum value of noise. It is clear that only 
a very narrow band of frequencies $d \nu$ around a third frequency $\approx 0.045$ 
(corresponding to the period $T \approx 22s$), are 
amplified at the output. Note that this frequency is absent in the input and 
corresponds to $(\nu_1 - \nu_2)/2$. This is in sharp contrast to the earlier 
case of a bistable system. The variation of escape probability corresponding to 
this frequency as a function of noise amplitude is shown in Fig.14.

This result can be understood as follows: When two signals of frequencies 
$\nu_{1}$ and $\nu_{2}$ and equal amplitude $Z$ are superposed, the resulting 
signal consists of peaks of amplitude $2Z$ repeating with a frequency 
$(\nu_{1} - \nu_{2})/2$ in accordance with the linear superposition 
principle:
\begin{equation}
Sin(2 \pi \nu_{1} t) + Sin(2 \pi \nu_{2} t) = 2Sin(2 \pi \nu_{+}t)Cos(2 \pi \nu_{-}t)
\label{eqn22}
\end{equation}
where $\nu_{+}=(\nu_{1}+\nu_{2})/2$ and $\nu_{-}=(\nu_{1}-\nu_{2})/2$. 
For a threshold system, the probability of escape depends only on the amplitude of 
the signal which is maximum corresponding to the frequency $\nu_{-}$. But 
in the case of a bistable system, the signal is enhanced only if there is a
regular shuttling between the wells at the corresponding frequency. This is 
rather difficult for the frequency $\nu_{-}$ because, its amplitude is 
modulated by a higher frequency $\nu_{+}$. This result has been checked by 
using different combinations of frequencies $(\nu_{1},\nu_{2})$ and also with 
different amplitudes. It should be mentioned here that this result reveals a 
fundamental difference between the two mechanisms of SR and is independant of 
the model considered here. We have obtained identical results with a 
fundamentally different model showing SR, namely, a model for Josephson 
junction and has been discussed elsewhere[21].

\begin{figure}
\resizebox{1.0\columnwidth}{!}{\includegraphics{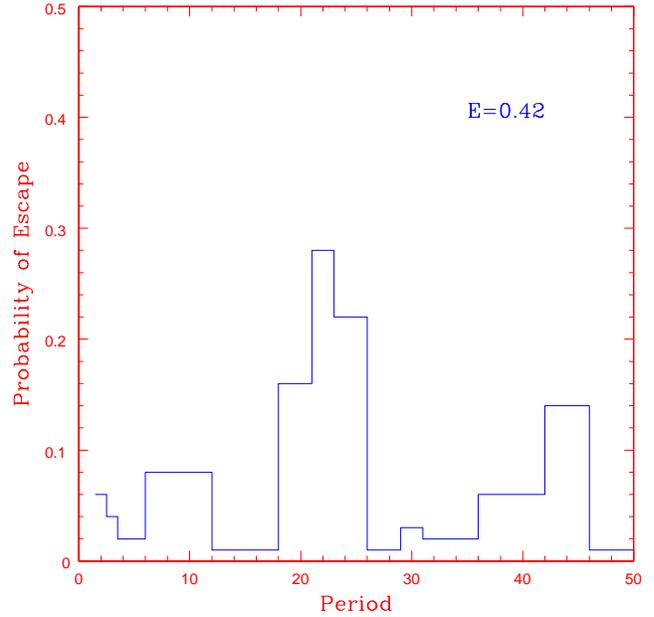}}
\caption{The escape probability corresponding to the optimum noise 
amplitude with the signal $F(n)$ comprising of 2 frequencies $\nu_1=0.125$ 
and $\nu_2=0.033$. The amplitudes of individual signals are put much below 
the threshold value required for SR. Note that only the difference frequency 
$(\nu_1 - \nu_2)/2$ is amplified.}
\label{Fig.13label}
\end{figure}

\begin{figure}
\resizebox{1.0\columnwidth}{!}{\includegraphics{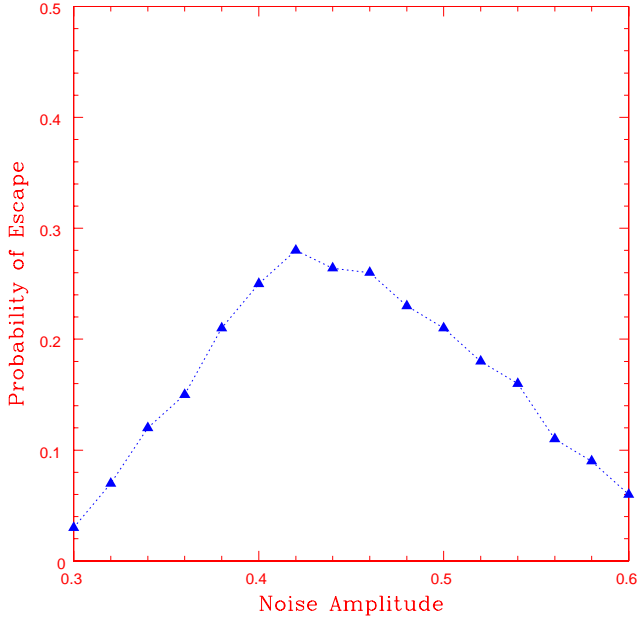}}
\caption{Variation of the probability of escape corresponding to the frequency 
shown in $Fig.13$, as a function of noise amplitude.}
\label{Fig.14label}
\end{figure}

To study VR in the threshold set up, we once again consider the input signal as in 
eq.(13) with the low frequency $\nu_1$ fixed at $0.125$ and $Z(1)=0.16$. With 
$E=0.12$ and $\nu_2=0.32$, the computations are performed as before by tuning the high 
frequency amplitude $Z(2)$. The results are shown in Fig.15, which clearly indicates 
VR in the threshold set up. The computations are repeated for different values of 
$\nu_2$ and here also it is found that the optimum escape probability becomes less 
for $\nu_2 > 0.45$ as in the bistable case.

It is interesting to note that when the high frequency forcing and noise are 
present simultaneously, only one of them dominates to enhance the signal. In the 
regime of VR, the noise level has to be small enough and vice versa for SR. When 
both are high, the transitions become random and the signal is lost. We find that 
the two regimes of (VR and SR) can be distinguished in terms of an initial time 
delay for the system to respond to the high frequency or noise, called the 
\emph{response time}. The response time $(\tau)$ decreases sharply in the regime of 
SR (ie, when signal is boosted with the help of noise), thus indicating a cross 
over behavior between the two regimes. This is shown in Fig.16, where $\tau$ is 
plotted as a function of $E$ for the threshold setup. The computations are 
done in such a way that as $E$ is increased, $Z(2)$ is decreased 
correspondingly to get optimum response.

\begin{figure}
\resizebox{1.0\columnwidth}{!}{\includegraphics{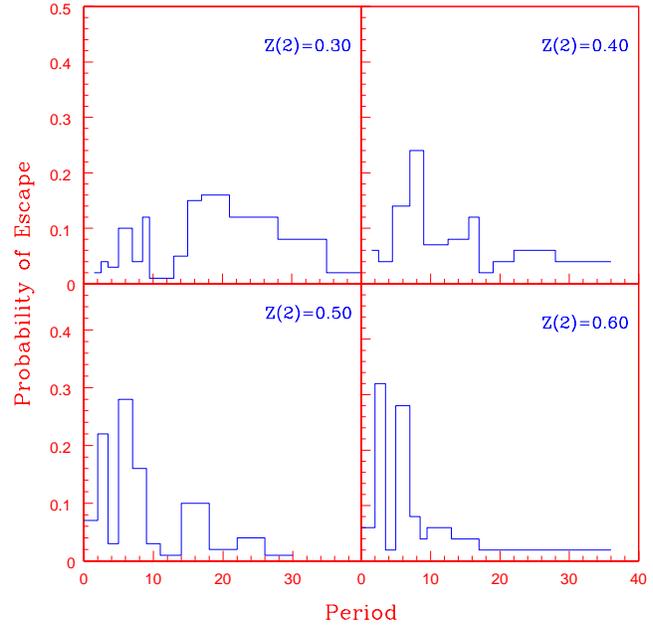}}
\caption{Escape probability as a function of different periods corresponding to 
four values of the signal amplitude of the high frequency $(\nu_2=0.32)$, showing VR 
for the low frequency signal $\nu_1$ for the cubic map in the threshold setup. 
Other parameters are $a=1.4, b=0.01$, $E=0.12$ and $Z(1)=0.16$.}
\label{Fig.15label}
\end{figure}

\begin{figure}
\resizebox{1.0\columnwidth}{!}{\includegraphics{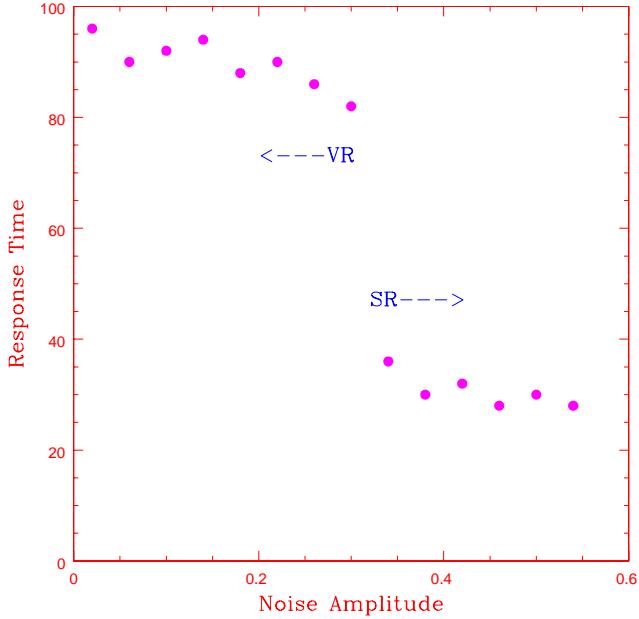}}
\caption{Variation of response time $\tau$ for VR and SR as a function of noise 
amplitude $E$, corresponding to the frequency $\nu_1 = 0.125$, for the cubic map 
in the threshold setup. For each $E$, the optimum value of the amplitude $Z(2)$  
for the higher frequency $\nu_2 = 0.32$ is used. Two regions for 
VR and SR are clearly evident in terms of the response time.}
\label{Fig.16label}
\end{figure}

\section{Results and discussion}
\label{sec:6}
In this paper we undertake a detailed numerical study of SR and VR in a discrete 
model with a bichromatic input signal and gaussian noise. The model combines the 
features of bistable and threshold setups and turns out to be ideal for comparing 
various aspects of SR and VR. Both additive and multiplicative noise are used 
for driving the system in the bistable mode. To our knowledge, this is the 
first instance where VR has been explicitely shown in a discrete system.

The system shows bistability in periodic and chaotic attractors. Hence a subthreshold 
signal can be enhanced with the inherent chaos in the system in the absence of 
noise. With a composite input signal and noise, all the component frequencies 
are enhanced in the output at different optimum noise amplitudes.

Our analysis brings out some fundamental differences between the two 
mechanisms of SR with respect to amplification of multisignal inputs. In particular,  
we find that, while the bistable set up responds only to the fundamental frequencies 
present in the input signal, the threshold mechanism enhances a mixed mode also, 
a result not possible in the context of linear signal processing. This may have 
potential practical applications, especially in the study of neuronal mechanisms 
underlying the detection of pitch of complex tones [19,36].

Another especially interesting result we have obtained is regarding a cross 
over behavior between VR and SR in terms of an initial response time. The 
response time decreases sharply in going from VR to SR and depends only on 
the background noise level. In the regime of VR, the system shows the 
so called bona fide resonance, where the response becomes optimum for a 
narrow band of intermediate frequencies. Moreover, the optimum value of 
high frequency signal amplitude increases with frequency, but as the frequency 
increases beyond a limit, it becomes less effective in producing VR.  

Finally, two other results obtained in the context of SR are the usefulness of a high 
frequency signal in enhancing the SNR of a low frequency signal and the possibility 
of using SR as a \emph {filter} for the detection or selective transmission of 
fundamental frequencies in a composite signal using a bistable nonlinear medium. 
The former is evident from Fig.8 and shows certain cooperative behavior 
between the two signals. Similar results have been obtained earlier [20,37] 
under other specific dynamical setups, but the simplicity of our model suggests 
that the results could be true in general. The latter result arises due to the 
fact that the noise amplitudes for the optimum SNR for the two frequencies are 
different (see Fig.5 and Fig.8) and the difference $\Delta E$ tend to increase 
with the difference in frequencies $(|\nu_1 - \nu_2|)$. This suggests that SR 
can, in principle, be used as an effective tool for signal detection/transmission 
in noisy environments. A similar idea has been proposed recently [18] in connection 
with the signal propogation along a  one dimensional chain of coupled 
over damped oscillators. There it was shown that noise can be used to 
select the harmonic components propogated with higher efficiency along the chain.

\begin{acknowledgement}
The authors thank the hospitality and computing facilities in IUCAA, Pune.
\end{acknowledgement}

\end{document}